\documentclass[final,5p, times,number]{elsarticle}
\usepackage[utf8]{inputenc}
\usepackage{amsmath}
\usepackage{amssymb}
\usepackage{booktabs}
\usepackage[frozencache,cachedir=./_minted]{minted}
\usepackage{multirow}
\usepackage{xcolor}
\usepackage{subcaption}
\usepackage{enumitem}
\usepackage{algorithm}
\usepackage{algpseudocode}
\usepackage{placeins}
\usepackage{tcolorbox}
\usepackage{tikz}
\usetikzlibrary{arrows.meta, positioning}      
\usepackage{alltt}
\tcbuselibrary{skins, breakable}
\usepackage{hyperref}

\setminted{
  frame=lines,
  breaklines=true,
  linenos=true,
  tabsize=2
}
\definecolor{rqbg}{HTML}{F3F3F3}      
\definecolor{rqtitle}{HTML}{444444}  
\definecolor{evintro}{HTML}{1B5E20}  
\definecolor{evchange}{HTML}{E65100} 
\definecolor{evremove}{HTML}{B71C1C} 
\definecolor{algcmt}{HTML}{757575}   
\definecolor{gitcmd}{HTML}{4527A0}   

\newtcolorbox{rqsummary}[2][]{%
  enhanced,
  breakable,
  colback=rqbg,
  colframe=white,
  boxrule=0pt,
  arc=0pt,
  left=6mm,
  right=2mm,
  top=2mm,
  bottom=2mm,
  before skip=10pt,
  after skip=10pt,
  overlay={
    \fill[rqtitle] (frame.south west) rectangle ([xshift=5mm]frame.north west);
    \node[rotate=90, anchor=center, text=white, font=\sffamily\small]
      at ([xshift=2.5mm]frame.west) {#2};
  },
  #1
}

\newcommand{\libs}{\textit{libraries \& frameworks}}

\newcommand{\others}{\textit{others}}
\newcommand{\lvlSimple}{\textit{Simple}}
\newcommand{\lvlOptional}{\textit{Optional}}
\newcommand{\lvlUnion}{\textit{Union}}
\newcommand{\lvlCollection}{\textit{Collection}}
\newcommand{\lvlNested}{\textit{Nested}}
\newcommand{\rqcTotalPeriods}{793{,}711}
\newcommand{\rqcIntroducedPct}{84.2}
\newcommand{\rqcChangedPct}{11.3}
\newcommand{\rqcRemovedPct}{4.5}

\newcommand{\rqcUniqueMembers}{632{,}404}
\newcommand{\rqcIntactN}{547{,}508}
\newcommand{\rqcIntactPct}{86.6}
\newcommand{\rqcModifiedN}{63{,}345}
\newcommand{\rqcModifiedPct}{10.0}
\newcommand{\rqcRemovedMemberN}{26{,}394}
\newcommand{\rqcRemovedMemberPct}{4.2}
\newcommand{\rqcParamPct}{52.34}
\newcommand{\rqcReturnPct}{27.25}
\newcommand{\rqcVarPct}{20.42}
\newcommand{\rqcUnionNetShare}{90.5}
\newcommand{\rqcOptionalRetention}{18.6}
\newcommand{\rqcOptionalToUnionPct}{44.7}
\newcommand{\rqcSimpleOutN}{9{,}645}
\newcommand{\rqcSimpleOutPct}{35.9}
\newcommand{\rqcSimpleToUnionN}{3{,}271}
\newcommand{\rqcSimpleToUnionPct}{12.2}
\newcommand{\rqcSimpleToCollectionN}{2{,}749}
\newcommand{\rqcSimpleToCollectionPct}{10.2}
\newcommand{\rqcSimpleToOptionalN}{2{,}999}
\newcommand{\rqcSimpleToOptionalPct}{11.2}
\newcommand{\rqdBothCount}{443{,}318}
\newcommand{\rqdSimpleAgree}{95.2}
\newcommand{\rqdSimpleToUnion}{3.0}
\newcommand{\rqdSimpleToCollection}{1.5}
\newcommand{\rqdSimpleToNested}{0.3}
\newcommand{\rqdOptionalToSimple}{21.5}
\newcommand{\rqdOptionalToUnion}{78.3}
\newcommand{\rqdOptionalToCollection}{0.2}
\newcommand{\rqdUnionToSimple}{29.9}
\newcommand{\rqdUnionAgree}{67.4}
\newcommand{\rqdUnionToCollection}{2.3}
\newcommand{\rqdUnionToNested}{0.3}
\newcommand{\rqdCollectionToSimple}{24.5}
\newcommand{\rqdCollectionToUnion}{5.1}
\newcommand{\rqdCollectionAgree}{67.2}
\newcommand{\rqdCollectionToNested}{3.1}
\newcommand{\rqdNestedToSimple}{24.5}
\newcommand{\rqdNestedToUnion}{41.9}
\newcommand{\rqdNestedToCollection}{9.1}
\newcommand{\rqdNestedAgree}{24.5}
\newcommand{\rqdTotalMembers}{3{,}660{,}483}
\newcommand{\rqdPyrOnly}{79.9}
\newcommand{\rqdDevOnly}{1.7}
\newcommand{\rqdBothVenn}{12.1}
\newcommand{\rqdNeither}{6.3}
\newcommand{\rqdUnannotatedInferable}{92.7}
\newcommand{\rqdParamBoth}{21.2}
\newcommand{\rqdParamDevOnly}{1.1}
\newcommand{\rqdParamPyrOnly}{74.0}
\newcommand{\rqdParamNeither}{3.6}
\newcommand{\rqdReturnBoth}{20.4}
\newcommand{\rqdReturnDevOnly}{2.5}
\newcommand{\rqdReturnPyrOnly}{66.8}
\newcommand{\rqdReturnNeither}{10.3}
\newcommand{\rqdVarBoth}{2.7}
\newcommand{\rqdVarDevOnly}{1.9}
\newcommand{\rqdVarPyrOnly}{88.6}
\newcommand{\rqdVarNeither}{6.8}
\newcommand{\rqa}{RQ1. To what extent are type hints adopted in Python libraries and frameworks?}
\newcommand{\rqb}{RQ2. How do library and framework developers use type hints?}
\newcommand{\rqc}{RQ3. How do developers introduce, change, and remove type hints in Python libraries and frameworks?}
\newcommand{\rqd}{RQ4. Why do developers introduce type hints in libraries and frameworks?}

\newcommand{\noindentparagraph}[1]{{\setlength{\parindent}{0pt} \setlength{\parskip}{0.5em} \textit{\textbf{#1}}\enspace}}

\makeatletter
\providecommand\@adddotafter[1]{#1\@addpunct{.}}
\makeatother

\algrenewcommand\algorithmiccomment[1]{\hfill\textcolor{algcmt}{$\triangleright$ #1}}  

\journal{Information and Software Technology}

\begin{document}

\begin{frontmatter}

  \title{Type Hints in Python Libraries and Frameworks: An Empirical Analysis of
    Adoption and Maintenance}


  \author[ufmg]{Thiago Roberto Magalh\~aes}
  \ead{trm@ufmg.br}

  \author[etsmtl]{Fabio Petrillo}
  \ead{fabio.petrillo@etsmtl.ca}

  \author[ufmg]{Jo\~ao Eduardo Montandon\corref{cor1}}
  \ead{joao@dcc.ufmg.br}

  \cortext[cor1]{Corresponding author}

  \affiliation[ufmg]{organization={Universidade Federal de Minas Gerais},
    city={Belo Horizonte},
    state={Minas Gerais},
    country={Brazil}}

  \affiliation[etsmtl]{organization={\'Ecole de Technologie Sup\'erieure},
    city={Montr\'eal},
    state={Qu\'ebec},
    country={Canada}}

  \begin{abstract}
    \textbf{Context:} In Python, type hints allow developers to annotate variables and functions with explicit type information, improving code clarity and reliability.
    Although type hints are widely available, little is known about how they are adopted and maintained in libraries and frameworks.
    \textbf{Objective:} We investigate the adoption, usage, maintenance, and rationale of type hints in Python libraries and frameworks.
    \textbf{Method:} We analyzed 1,000 popular GitHub repositories, identifying libraries and frameworks and extracting their type annotations.
    We examined annotation coverage, the locations and origins of annotations, their evolution across git histories, and the relationship between developer annotations and types inferred by Pyright.
    \textbf{Results:} Of the analyzed repositories, 91\% of libraries use type hints at least once, although adoption is inconsistent.
    Among libraries with systematic usage, maintainers prioritize function parameters and return types, with median coverage of 45.8\% and 35.9\%, respectively, and mainly use built-in types (73.0\%).
    When modified, annotations tend to migrate to more expressive types.
    Developers annotate members even when Pyright can infer their types, and these annotations often simplify the inferred type.
    \textbf{Conclusion:} Type hints in Python libraries and frameworks primarily serve as API contracts rather than comprehensive descriptions of implementation details.
    The findings suggest opportunities for tooling that prioritizes public interfaces, identifies meaningful annotation changes, and supports maintainers in evolving type information.
  \end{abstract}



\end{frontmatter}

\section{Introduction} 

Libraries and frameworks are the building blocks of modern software development~\cite{Montandon2019, Terragni2024, Lamothe2021}.
These components provide pre-built functionalities that speed up the implementation of software products by allowing developers to focus on core business logic.
This modular approach not only enhances code reusability but also reduces maintenance costs, allowing teams to deliver robust and scalable solutions faster and cheaper.
In this context, the Python ecosystem emerges as one of the most active~\cite{Verbina2025, StackOverflow2024}, offering solutions for a wide range of software applications, from web development to data science and machine learning~\cite{Montandon2025, Zhang2021, Wang2020}.

Python was designed to be simple and clean, making it easy to read and write code~\cite{Scarlett2023, Ramalho2022}.
The language is dynamically typed, i.e., variable types are determined and checked at runtime rather than at compile time~\cite{Webber2002}.
Such features give developers a large degree of freedom when writing code, since they can handle different objects uniquely based on their behaviour.\footnote{\url{https://en.wikipedia.org/wiki/Duck\_typing}}
In practice, developers can implement programs faster and with less boilerplate~\cite{Mir2021,Webber2002}.
However, dynamic typing can also lead to potential issues, such as runtime type errors that are only detected during execution~\cite{Bogner2022,Du2022,DiGrazia2022,Guo2024, Webber2002}.
To aid static analysis tools in detecting type inconsistencies before runtime, Python introduced type hints in 2014 at version 3.5~\cite{Ramalho2022}.
This feature allows developers to annotate variables and functions with explicit type information, thus providing more context about which types are expected when using these annotated members.

Type hints have gained significant traction within the Python community since their introduction~\cite{Verbina2025}, being adopted in popular projects such \textit{NumPy}\footnote{\url{https://numpy.org/}} and \textit{Pandas}.\footnote{\url{https://pandas.pydata.org/}}
Prior work in the literature has shown the benefits of using types to improve error detection and software maintainability~\cite{Hanenberg2014,Gao2017,Bogner2022,Mezzetti2018,DiGrazia2022,Guo2024,Mir2021, Jin2021, Lin2023}.
Yet, little is known about how type hints are adopted in libraries and frameworks.

\noindentparagraph{Our Study.} By extending our previous work~\cite{Magalhaes2026}, we provide a more comprehensive understanding of how type hints are adopted, used, and maintained in real-world Python libraries and frameworks.
We collected and extracted type annotation data from 1,000 popular Python repositories hosted on GitHub, identified which ones are libraries and frameworks, and analyzed their type hint usage.
Specifically, we investigated the following research questions:

\begin{itemize}[itemsep=0.20cm, after=\vspace{0.20cm}, before=\vspace{0.20cm}]
  \item \textit{\rqa}
        Despite 91\% of libraries having used type hints at least once, this adoption is not employed consistently across their codebase.
        In fact, half of the libraries covered only 13.6\% of their members with types.
  \item \textit{\rqb}
        Considering libraries with systematic type hint usage, maintainers focus on annotating function parameters and return types, with median coverage of 45.8\% and 35.9\%, respectively.
        Furthermore, built-in types are preferred, accounting for 73.0\% of all type annotations in half of the libraries.
  \item \textit{\rqc}
        Across \rqcTotalPeriods{} annotation periods, \rqcIntroducedPct\% of events are introductions, \rqcChangedPct\% are changes, and only \rqcRemovedPct\% are removals.
        Type changes follow a clear directional trend: \lvlUnion{} is the dominant destination, receiving 3.3$\times$ more cross-level inflows than outflows, while \lvlOptional{} is the largest net donor.
        At the commit level, when both parameter and variable annotations change together, the return type also changes in over 60\% of those commits.
  \item \textit{\rqd}
        Comparing developer annotations against Pyright's static inference across \rqdTotalMembers{} members reveals that \rqdPyrOnly\% are covered only by Pyright, \rqdDevOnly\% only by developers, \rqdBothVenn\% by both, and \rqdNeither\% by neither. Developers frequently override Pyright's inferences, especially for parameters and return types, and systematically prefer \texttt{Optional}-based syntax.
\end{itemize}

We highlight the following contributions of this paper:
(1)~an empirical quantification of type hint adoption in Python libraries and frameworks;
(2)~an analysis of how type hints are maintained and evolve over time, showing that Union types dominate annotation changes;
(3)~evidence that developers annotate where static type inference is insufficient, offering insights into the complementary role of type hints and automated inference in gradual typing systems.

The remainder of this paper is organized as follows.
Section \ref{sec:background} introduces the concept of Type Hints.
Section \ref{sec:study-design} explains the methodology used to collect and analyze type hints data from Python repositories.
Our results are presented in Section \ref{sec:results}, whereas the implications of this work are discussed in Section \ref{sec:discussion}.
Sections \ref{sec:threats} and \ref{sec:related-work} present the threats to validity and related work.
Finally, Section \ref{sec:conclusion} concludes the paper and outlines future work.

\section{Understanding Type Hints} 
\label{sec:background}



Type Hints are one of the biggest changes in Python history~\cite{Ramalho2022}.
Introduced in PEP 484\footnote{\url{https://peps.python.org/pep-0484/}} in 2014, this proposal specifies the syntax and semantics for explicit type declarations in function arguments, return values, and variables.
Type hints are provided as a \textit{gradual type system}, where type annotations
(a) are optional, since Python checker should not emit a warning for code without types;
(b) do not prevent inconsistent values from being assigned during runtime, and
(c) do not improve the program's performance since types are not added to the program's bytecode.
\textit{The goal is to provide more type information to static analysis tools so they can detect inconsistencies more effectively before runtime}.

\subsubsection*{\textbf{A Type Hint Example}}

Figure \ref{fig:type-hint-example} presents an example of the use of type hints.
Lines 1 and 2 define the \mintinline{python}{concat} function, which receives two strings as parameters---\mintinline{python}{first} and \mintinline{python}{second}---and prints the concatenation of both on the screen.
Note that both parameters are annotated with the type \mintinline{python}{str}, the string type in Python, and the return value is annotated with \mintinline{python}{None}, representing a method without a return value.

\begin{figure}[htbp]
  \begin{minted}[
      frame=lines,
      framesep=2mm,
      breaklines,
      autogobble,
      linenos,
      fontsize=\small
    ]{python}
    def concat(first: str, second: str) -> None:
        print(first + second)

    concat("Type", "Hint")
    concat(1, 2)
  \end{minted}
  \caption{A function with type hints in Python.}
  \label{fig:type-hint-example}
\end{figure}

Line 4 calls \mintinline{python}{concat} with \mintinline{python}{"Type"} and \mintinline{python}{"Hint"} as parameters, which prints \mintinline{python}{"TypeHint"}.
This call is not only valid but also expected, since the type of the values involved in the call match the types annotated in \mintinline{python}{concat}'s definition.
Static analysis tools emit no warning on this call.

\begin{figure*}
  \centering
  \includegraphics[width=1\linewidth]{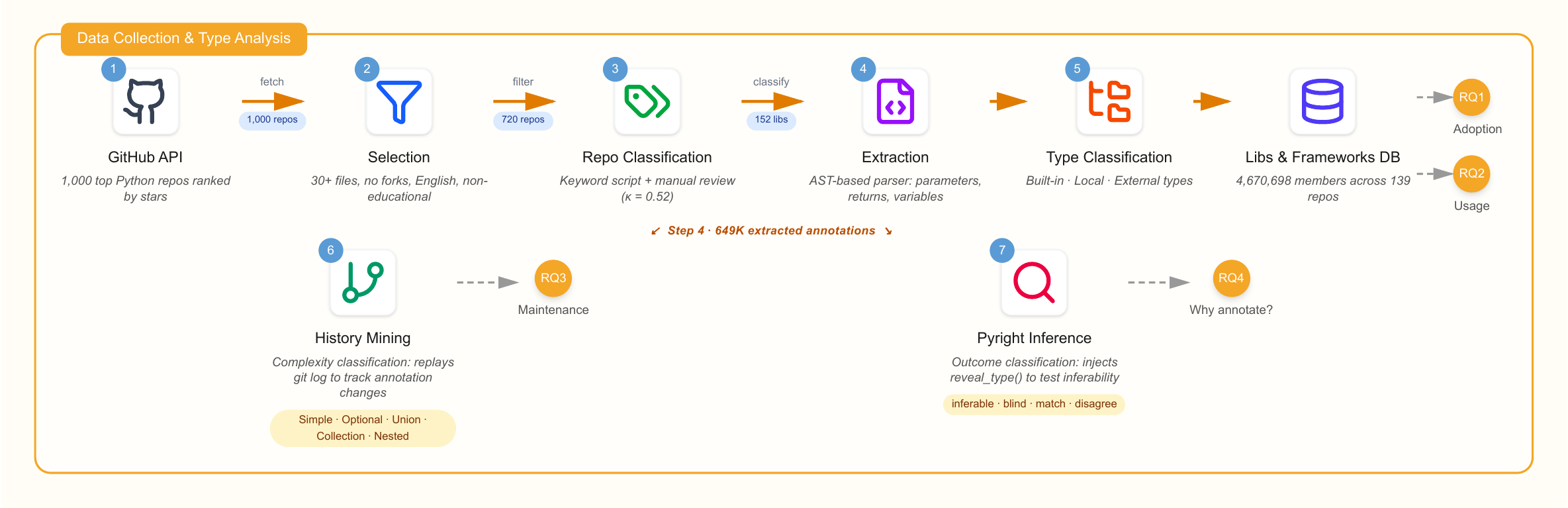}
  \caption{Data Collection adopted in this study.}
  \label{fig:diagram}
\end{figure*}

On the other hand, line 5 invokes the function with \mintinline{python}{1} and \mintinline{python}{2} as parameters.
This call clearly violates the annotated types, since both values belong to the \mintinline{python}{int} type.
Any static analysis tool shall emit a warning on this line, showing that the types are incompatible.
Nonetheless, this code will still execute and print \mintinline{python}{3} as a result, i.e., it will sum both \mintinline{python}{int} values.
This happens because, in a gradual type system, type hints do not prevent runtime type errors; they are intended to aid static analysis and tooling.


\section{Study Design} 
\label{sec:study-design}

We advocate the usage of type hints is particularly important in \libs, where the lack of clarity, correctness, and the presence of errors impact the reliability of the features provided by them~\cite{Mezzetti2018, Montandon2025, Jayasuriya2024,Du2022}.
This study provides a comprehensive assessment of how type hints are being used in Python third-party components.

\subsection{Research Questions}

In this study, we elaborated four research questions to be answered.

\paragraph{\rqa}
Prior work shows that the adoption of type hints is becoming more popular on Python projects~\cite{DiGrazia2022,Mir2021}.
However, we lack empirical evidence on how this feature is being adopted by \libs.
Thus, we first verify whether projects representing third-party components do use type hints in their codebase.

\paragraph{\rqb}
We conduct an exploratory investigation to understand how maintainers utilize type hints in \libs.
Particularly, we analyzed \underline{\textit{where}} type hints are mostly employed, and \underline{\textit{what}} are the most frequent annotated types.

\paragraph{\rqc}
To understand \underline{when} type hints are modified in \libs, we trace the annotation history of each annotated member across the git commit history. We leveraged the modification history behind each type hint and analyzed in which context these types are modified over time.


\paragraph{\rqd}
We investigate \underline{why} type hints are introduced in \libs\ by testing the hypothesis that they are added to overcome limitations of current static type checkers, such as Pyright.

\subsection{Data Collection}

The study design and methodology adopted in this work are depicted in Figure \ref{fig:diagram}, and are described as follows.

\subsubsection*{\textbf{1. Fetch GitHub Repositories}}
\label{step:1}

We initially gathered a collection of 1,000 Python repositories using the GitHub API, ranked by the number of stars.
This approach ensured our dataset contains well-established and widely adopted projects in the Python ecosystem from a diverse range of domains such as web development, data science, and machine learning.
We recorded repository metadata---such as repository name, number of stars, and number of Python files---for later reference.

\subsubsection*{\textbf{2. Repository Selection}}
\label{step:2}

To improve data quality and reduce noise, we applied several filtering criteria.
First, repositories containing fewer than 30 Python source files were excluded, since small projects rarely contain sufficient code to provide meaningful insights into annotation practices. This threshold was established as a pragmatic lower bound informed by manual inspection of the smallest repositories in our initial dataset: projects below this size tended to consist of a single module or a handful of scripts, offering too little surface area to draw meaningful conclusions about annotation discipline.
Second, we removed repositories primarily designed for educational purposes (e.g., tutorials, course materials, or example snippets), as they tend to exhibit artificially simplified code~\cite{icsme2016,Santana2024}. To support both this step and the subsequent classification step, we developed a script that scanned repository names, descriptions, and tags for two complementary sets of keywords: (a) \textit{library and framework indicators} (e.g., \textit{library}, \textit{framework}, \textit{sdk}, \textit{toolkit}, \textit{client}, \textit{api}) to pre-classify repositories as \libs; and (b) \textit{educational indicators} (e.g., \textit{tutorial}, \textit{course}, \textit{exercises}, \textit{bootcamp}, \textit{example}) to flag repositories for exclusion. Repositories for which neither set of keywords was detected were subject to manual inspection in the subsequent classification step.
Third, we excluded repositories that were forks or duplicates of other projects in our dataset to avoid redundancy and ensure diversity in our analysis.
Finally, we disregarded those whose primary language was not English.
After this step, 720 repositories were selected.

\subsubsection*{\textbf{3. Repository Classification.}}
\label{step:3}

Repositories for which the keyword script from \hyperref[step:2]{Step~2} produced a clear signal were assigned to their category automatically: those matching library and framework indicators were pre-classified as \libs, and those matching educational indicators were flagged for exclusion.
Only the remaining repositories, those for which neither set of keywords yielded a definitive result, required human judgment.
For these ambiguous cases, one author manually analyzed the name, description, and tags to decide whether each should be classified as \libs\ or \others~\cite{icsme2016}.
To assess inter-rater agreement on this manual step, a second author independently categorized a sample of 100 repositories.
The resulting Cohen's Kappa was 0.52, indicating moderate agreement between both classifications~\cite{Warrens2015}.
Differences between the two classifications were discussed until consensus was reached.
Each repository classified as \libs\ was cloned locally to enable further static analysis of its source code.

\subsubsection*{\textbf{4. Type Hints Extraction.}}
\label{step:4}

We performed an in-depth static analysis to identify and extract the type hints implemented in their codebase.
For this, we built a custom analyzer in Python based on the Abstract Syntax Tree (AST) module.\footnote{\url{https://docs.python.org/3/library/ast.html}}
For each repository, we parsed every Python file, generated their \textit{ast}, and visited the following nodes to extract their type hints.

\begin{itemize}
  \item \textit{Annotated assignments} (\texttt{AnnAssign}), to detect type hints applied to variables, regardless of their declaration scope.
  \item \textit{Function definitions} (\texttt{FunctionDef}), to extract both parameter and return type annotations.
\end{itemize}

\subsubsection*{\textbf{5. Type Hints Classification.}}
\label{step:5}

We classify the type hints into one of the following three categories according to their origin.

\begin{itemize}
  \item \textit{Built-in types}, representing types that are provided by the Python language, such as \mintinline{python}{int}, \mintinline{python}{str}, and \mintinline{python}{dict}.
        These types were mapped based on the list of built-in types in the official Python documentation.\footnote{\url{https://docs.python.org/3/library/stdtypes.html}}
  \item \textit{Local types}, corresponding to types declared locally in the repository, such as classes, data structures, or aliases.
        To make this verification possible, we leveraged all type declarations performed in the project where the annotated type was found and verified whether the annotated type matched any of these locally declared classes.
  \item \textit{External types}, referring to types that do not belong to the previous categories, thus originating from third-party libraries like \mintinline{python}{Tensor} from \textit{pytorch}, and \mintinline{python}{DataFrame} from \textit{pandas}.
\end{itemize}

Before applying this classification, we performed a normalization step to ensure consistent comparison across all extracted types.
We (a) removed any package prefixes, e.g.,~converting \texttt{typing.Dict} to \texttt{Dict}, and (b) unified equivalent constructs, e.g.,~making \texttt{Dict} $\equiv$ \texttt{dict}.
To avoid excessive fragmentation, we treated complex annotations conservatively:
for generics, only the container type was considered;
for union and optional types, we decomposed them into their base components, excluding \texttt{None}.

\subsubsection*{\textbf{6. Mining Type Hints Modification}}
\label{step:6}

To trace down how type hints are maintained over time, we analyzed the git commit history of each repository.
For each annotated member, we obtained its file commit history and re-applied our type hint extraction process to each snapshot.
This procedure is explained in Algorithm~\ref{alg:annotation_history}.

\begin{algorithm}[!ht]
  \renewcommand{\baselinestretch}{1}\small
  \caption{Trace Annotation History}
  \label{alg:annotation_history}
  \begin{algorithmic}[1]
    \Require $\textit{members\_by\_file}$: map $\text{file} \to [(member,\; type)]$ \Comment{grouped by file: one git traversal per file}
    \For{each $(file,\; members)$ \textbf{in} $\textit{members\_by\_file}$}
    \State $commits \leftarrow$ \textcolor{gitcmd}{\texttt{git log --reverse -- }}$file$ \Comment{oldest $\to$ newest}
    \State $last \leftarrow \{(m,t) \mapsto \mathbf{None} \mid (m,t) \in members\}$ \Comment{i.e.,~not yet annotated}
    \Statex
    \For{each $hash$ \textbf{in} $commits$}
    \State $src \leftarrow$ \textcolor{gitcmd}{\texttt{git show}} $hash\texttt{:}file$ \Comment{historical snapshot}
    \State $ann \leftarrow \Call{ASTExtract}{src}$ \Comment{Step~4 extractor}

    \For{each $(m,\; t)$ \textbf{in} $members$}
    \State $prev \leftarrow last[m,t]$                          \Comment{previous commit}
    \State $cur  \leftarrow ann.get((m,\;t),\; \mathbf{None})$ \Comment{current commit}
    \Statex
    \If{$cur = \mathbf{None}$ \textbf{and} $prev \neq \mathbf{None}$}
    \State emit \textcolor{evremove}{\textsc{removed}}
    \ElsIf{$cur \neq \mathbf{None}$ \textbf{and} $prev = \mathbf{None}$}
    \State emit \textcolor{evintro}{\textsc{introduced}}
    \ElsIf{$cur \neq prev$}
    \State emit \textcolor{evchange}{\textsc{changed}}
    \EndIf
    \State $last[m,t] \leftarrow cur$ \Comment{update state for next commit}
    \EndFor
    \EndFor
    \EndFor
  \end{algorithmic}
  \vspace{0.4em}
  {\footnotesize
    \textbf{Notation.}
    $\textit{commits}$: list of commit hashes retrieved oldest-first from \texttt{git log}; commit date and message are stored as output metadata but do not affect the transition logic.
    $\textit{last}$: map from $(m,t)$ to the annotation at the previous commit (\textbf{None} = not yet seen).
    $\textit{prev}$: value of $last[m,t]$ at the start of the current iteration, i.e.,~the annotation before this commit.
    $\textit{src}$: raw file content at the given commit.
    $\textit{ann}$: map from $(m,t)$ to annotation string extracted by \textsc{ASTExtract}.
    $\textit{cur}$: annotation value at the current commit (\textbf{None} if the member is unannotated).

  }
\end{algorithm}

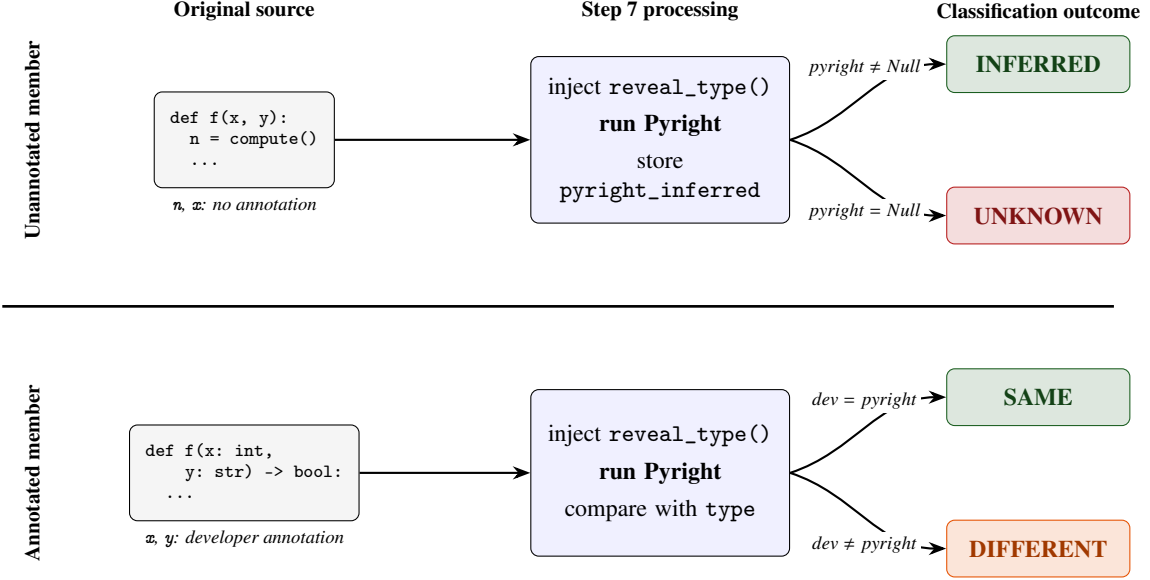
\begin{figure*}[ht]
  \centering
  \begin{tikzpicture}[
      font=\small,
      box/.style={draw, rounded corners=3pt, inner sep=6pt, align=center},
      code/.style={box, fill=gray!8, align=left, font=\ttfamily\scriptsize},
      proc/.style={box, fill=blue!6, text width=3.0cm, minimum height=2.2cm},
      result/.style={box, font=\bfseries\small, text width=2.0cm, minimum height=0.75cm},
      arr/.style={-Stealth, thick},
      lbl/.style={font=\scriptsize\itshape, fill=white, inner sep=1.5pt},
      hdr/.style={font=\footnotesize\bfseries, align=center},
    ]

    \node[code] (src_u) at (0, 2.2) {%
      def~f(x,~y):\strut\\
      \quad n~=~compute()\strut\\
      \quad ...\strut
    };
    \node[lbl, below=2pt of src_u] {\texttt{n}, \texttt{x}: no annotation};

    \node[proc] (pyr_u) at (5.5, 2.2) {%
      inject \texttt{reveal\_type()}\\[3pt]
      \textbf{run Pyright}\\[3pt]
      store \texttt{pyright\_inferred}
    };

    \node[result, fill=evintro!15, draw=evintro, text=evintro!70!black]
    (inf)   at (10.5,  3.2) {INFERRED};
    \node[result, fill=evremove!15, draw=evremove, text=evremove!70!black]
    (blind) at (10.5,  1.2) {UNKNOWN};

    \draw[arr] (src_u) -- (pyr_u);
    \draw[arr] (pyr_u.east) to[out=18, in=180] (inf.west);
    \draw[arr] (pyr_u.east) to[out=-18, in=180] (blind.west);
    \node[lbl] at (8.2, 3.15) {$pyright \neq Null$};
    \node[lbl] at (8.2, 1.25) {$pyright = Null$};

    \node[code] (src_a) at (0, -2.2) {%
      def~f(x:~int,\strut\\
      \quad\quad y:~str)~->~bool:\strut\\
      \quad ...\strut
    };
    \node[lbl, below=2pt of src_a] {\texttt{x}, \texttt{y}: developer annotation};

    \node[proc] (pyr_a) at (5.5, -2.2) {%
      inject \texttt{reveal\_type()}\\[3pt]
      \textbf{run Pyright}\\[3pt]
      compare with \texttt{type}
    };

    \node[result, fill=evintro!15, draw=evintro, text=evintro!70!black]
    (match) at (10.5, -1.2) {SAME};
    \node[result, fill=evchange!15, draw=evchange, text=evchange!70!black]
    (dis)   at (10.5, -3.2) {DIFFERENT};

    \draw[arr] (src_a) -- (pyr_a);
    \draw[arr] (pyr_a.east) to[out=18, in=180] (match.west);
    \draw[arr] (pyr_a.east) to[out=-18, in=180] (dis.west);
    \node[lbl] at (8.2, -1.25) {$dev = pyright$};
    \node[lbl] at (8.2, -3.15) {$dev \neq pyright$};

    \node[hdr] at (0,   3.9) {Original source};
    \node[hdr] at (5.5, 3.9) {Step 7 processing};
    \node[hdr] at (10.5, 3.9) {Classification outcome};

    \node[hdr, rotate=90] at (-2.8,  2.2) {Unannotated member};
    \node[hdr, rotate=90] at (-2.8, -2.2) {Annotated member};
    \tikzset{separator/.style={draw=black, line width=1pt}}

    \draw[separator] (-3.2, 0) -- (11.5, 0);

  \end{tikzpicture}
  \caption{Illustration of the \hyperref[step:7]{Step~7} procedure (Detecting Inferred Types).
    \texttt{reveal\_type()} probes are injected into each file, with only the target member's annotation removed so Pyright infers independently while retaining surrounding type context.
    Each member's stored pair (\texttt{type}, \texttt{pyright\_inferred}) is then classified into one of four outcomes.}
  \label{fig:rq4_procedure}
\end{figure*}

For each file containing annotated members, we retrieve the ordered list of commits that touched it (lines 2--3).
Then, for each commit, we fetch the file snapshot at that point in history and parse it with our AST-based extractor to obtain the set of annotations present at that revision (lines 6--7).
Finally, for each tracked member, we compare the annotation at the current commit ($cur$, line~10) against the annotation from the previous commit ($prev$, line~9): if $cur$ is absent while $prev$ was present, we record a \textsc{removed} event; if $cur$ is present while $prev$ was absent, we record an \textsc{introduced} event; and if both are present but differ, we record a \textsc{changed} event (lines 11--16).
After each comparison, $last[m,t]$ is updated with $cur$ so the next commit sees the current state as its $prev$ (line~18).

By comparing consecutive snapshots, we detected three types of annotation events: (a)~\textit{introduced}, when a member receives an annotation for the first time; (b)~\textit{changed}, when the annotation text differs from the previous commit; and (c)~\textit{removed}, when a previously annotated member loses its annotation. The annotation events collected in this step allow us to reconstruct a maintenance timeline for each type hint, which is used to answer RQ3.


To analyze how annotations evolve structurally, we classify each annotation into one of five \textit{complexity levels}. Unlike the origin-based taxonomy used in RQ2 (built-in, local, external, which captures \textit{what} a type refers to, the complexity classification captures \textit{how refined} an annotation is.
This way, we can investigate if a developer may adjust \mintinline{python}{int} into \mintinline{python}{Optional[int]}, or \mintinline{python}{list} into \mintinline{python}{List[str]}, regardless of the type's origin.
\begin{enumerate}
  \item \lvlSimple{} --- bare types with no parameterization, regardless of origin (e.g.,~\mintinline{python}{str}, \mintinline{python}{int}, or a local type class \mintinline{python}{MyModel});
  \item \lvlOptional{} --- \mintinline{python}{Optional[T]} where \texttt{T} is itself non-generic;
  \item \lvlUnion{} --- \mintinline{python}{Union[...]} or pipe-notation types (e.g.,~\mintinline{python}{str | int});
  \item \lvlCollection{} --- parameterized containers with one bracket level (e.g.,~\mintinline{python}{List[str]}, \mintinline{python}{Dict[str, int]});
  \item \lvlNested{} --- deeply nested collections or \mintinline{python}{Optional} over a collection type (e.g.,~\mintinline{python}{Optional[List[str]]}, \mintinline{python}{Dict[str, List[int]]}).
\end{enumerate}

Note that this taxonomy is orthogonal to the origin-based one: a local type such as \mintinline{python}{MyModel} is classified as \lvlSimple{} when used bare, \lvlOptional{} inside \mintinline{python}{Optional[MyModel]}, and \lvlCollection{} inside \mintinline{python}{List[MyModel]}.


\subsubsection*{\textbf{7. Detecting Inferred Types}}
\label{step:7}

As stated in Section~\ref{sec:background}, a gradual type system aims to feed static analysis tools with type information so they detect inconsistencies before runtime.
Thus, we hypothesize that type hints are used to give context that static type checkers are unable to solve otherwise.
We designed a comparative evaluation to verify this claim.

For each member, we created an alternative copy of its source file ensuring only that member's type information is unavailable while keeping all other type annotations intact as context; i.e.,~for annotated members we remove their type hints, and unannotated ones remain as is.
Next, we ran Pyright---a widely used static type checker for Python---in inference mode.
This mode predicts types for all members (unannotated members directly, and previously-annotated members after their annotation removal) based on the context provided during this execution.
We compared the type originally provided, if any, against the one inferred by Pyright to determine whether they match or differ.

Four possible outcomes arise from this comparison, as illustrated in Figure~\ref{fig:rq4_procedure}:
(a)~\textit{inferred}, when Pyright infers a type for an unannotated member;
(b)~\textit{unknown}, when Pyright was not able to find a type;
(c)~\textit{same}, when both developer and Pyright annotate the same type; and
(d)~\textit{different}, when the types provided by developer and Pyright differ.

\subsection{The Resulting Dataset}

From the 720 repositories selected for our study, 152 (~21\%) are \libs.
For each type hint detected, we stored: (a) the file path, (b) the qualified member name, (c) the annotated type, (d) the member category (local variable, parameter, or return), and (e) a contextual code snippet with the extracted annotation.
In total, this procedure analyzed 5,030,649 source code members and extracted 649,099 type hints.
We used this baseline to answer the research questions proposed for this work.

\paragraph{\textbf{Type Hint Coverage Metric}}
To properly assess the adoption of type hints across \libs, we computed a proportion-based metric called \textit{type hint coverage}.
For a given repository, this metric calculates the ratio of annotated members to the number of eligible members in the context under analysis.
For example, the \textit{flask} framework contains 4,614 overall members---i.e.,~parameters, return commands, and variables---of which 1,011 are annotated with type hints.
Thus, its \textit{overall type hint coverage} is 21.9\% ($1,011 / 4,614$).
From these, 1,808 are parameters and 524 of them are annotated, resulting in a \textit{parameter type hint coverage} of 28.9\% ($524 / 1,808$).
We used this metric as it allows fair comparison between repositories of different sizes and characteristics.

\section{Results} 
\label{sec:results}

\subsubsection*{\textbf{\rqa}}

Overall, 139 out of 152 (91\%) \libs\ contain at least one type hint in their codebase.
The project with the highest type hint coverage is \textit{openai-python}---the official library for the OpenAI API---with 61\% of its members with type hints, followed by \textit{altair}--- a library for data visualization---with 53\%.
Other popular projects show consistent adoption across their codebases.
\textit{Pytest}---one of the most popular Python libraries for software testing---annotates 39\% of its members.
\textit{FastAPI}---a well-known web development framework---fulfilled 35\% of its members with type hints.

\begin{figure}[htbp]
  \centering
  \includegraphics[width=0.9\linewidth]{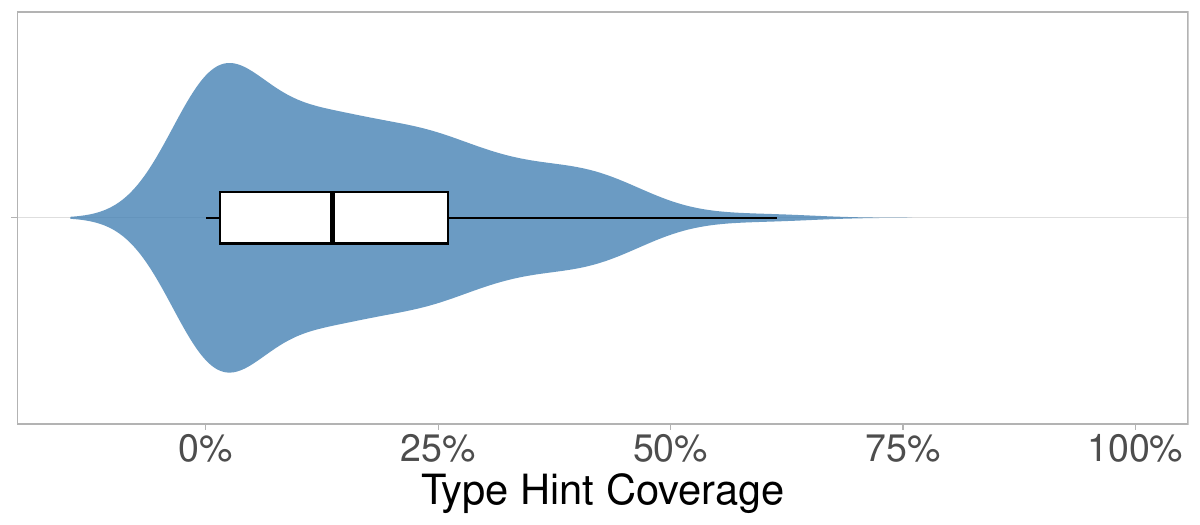}
  \caption{Distribution of overall type hint coverage.}
  \label{fig:rq1-overall}
\end{figure}

Despite these successful cases, most projects do not apply type hints consistently across their codebase.
Figure \ref{fig:rq1-overall} shows the distribution of type hint coverage among \libs.
The median type hint coverage is 13.6\%, while the third quartile annotated 26\% of their eligible members.

\begin{rqsummary}{Finding \#1}
  Overall, 91\% of \libs\ include at least one type hint, which does not mean type hints are consistently used.
  On the contrary, half of the \libs\ reported 13\% of type hint coverage.
\end{rqsummary}

\subsubsection*{\textbf{\rqb}}

\begin{figure}[ht]
  \begin{subfigure}[b]{1\linewidth}
    \centering
    \includegraphics[width=0.91\linewidth]{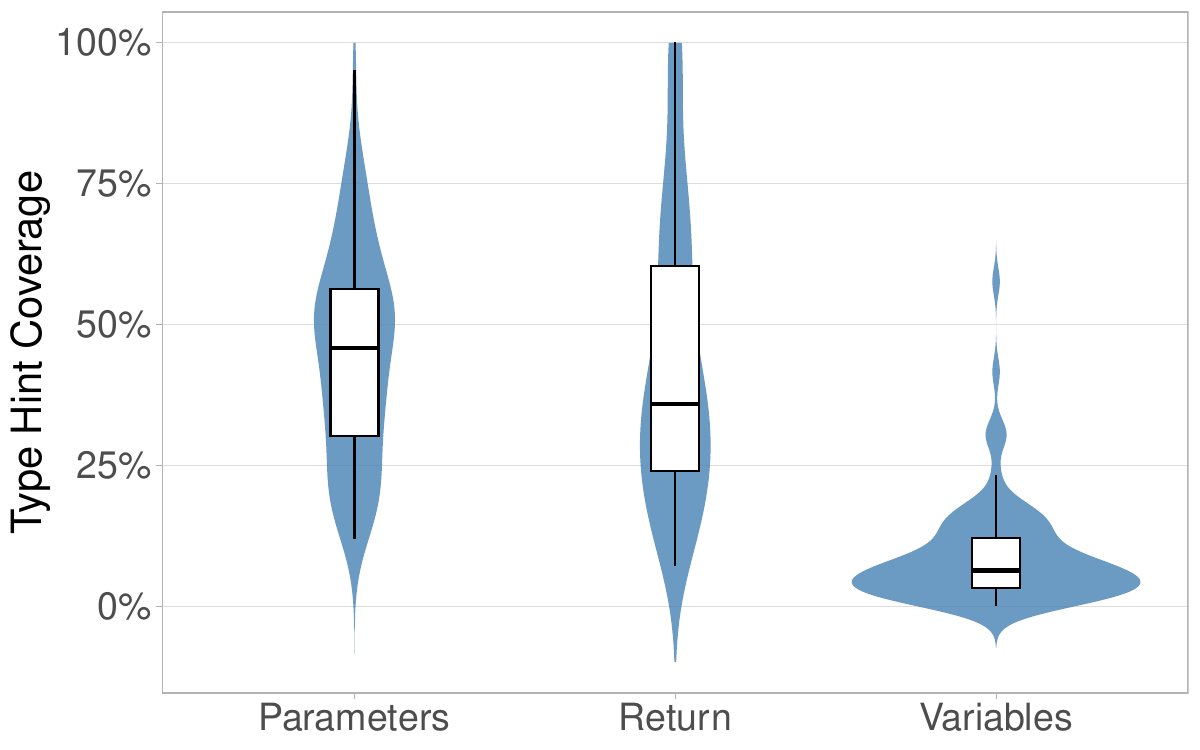}
    \caption{Type Hint Location}
    \label{fig:rq2-location}
  \end{subfigure}
  \begin{subfigure}[b]{1\linewidth}
    \centering
    \includegraphics[width=0.91\linewidth]{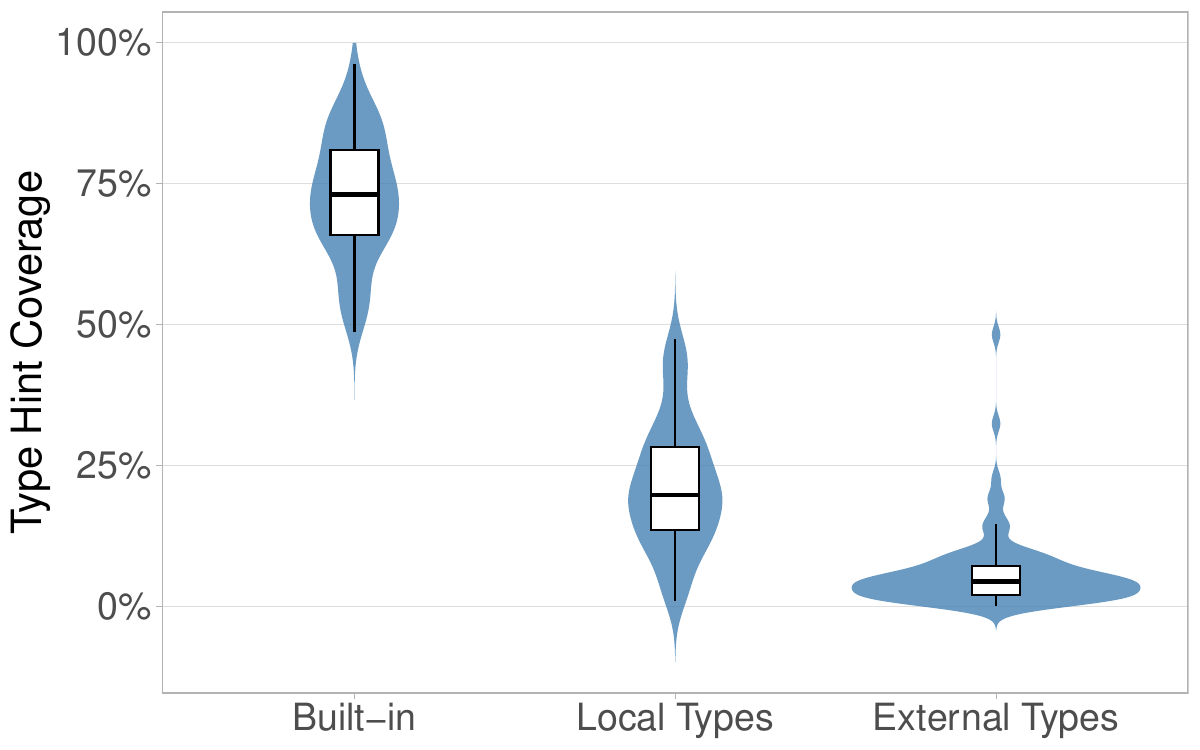}
    \caption{Type Hint Level}
    \label{fig:rq2-level}
  \end{subfigure}
  \caption{Distribution of type hint usage according to their location and type levels.}
  \label{fig:rq2}
\end{figure}

To address RQ2, we focused on \libs\ with at least 10\% overall type hint coverage, filtering out those with occasional usage, resulting in a subset of 86 repositories.
Figure \ref{fig:rq2} presents the distribution of the type hint coverage according to \textit{where} they were applied, and \textit{what} types were used in the annotations.

The distribution of type hint usage according to their location is shown in Figure \ref{fig:rq2-location}.
As we can see, parameters and return types are significantly more annotated than variables.
The median type hint coverage for parameters and return types is 45.8\% and 35.9\%, respectively, while variables present a median coverage of only 6.3\%.
We performed a Spearman's correlation analysis to assess the relationship between the coverage of these code members~\cite{Parnin2013, Sprent2016}.
We found a strong correlation between the coverage of parameters and return types ($\rho = 0.64$, $p < 0.05$); by contrast, variable coverage shows a weak correlation with both parameters ($\rho = 0.27$, $p < 0.05$) and return types ($\rho = 0.28$, $p < 0.05$).
This suggests that library maintainers prioritize annotating function signatures over local implementation details, probably to enhance API clarity and increase library documentation.

Figure \ref{fig:rq2-level} illustrates the distribution of type hint coverage according to their type levels.
Built-in types are largely adopted, as 50\% of the \libs\ present 73.0\% of their type hints using built-in types.
On the other hand, local object types are less frequently used, with a median coverage of 19.7\%.
External object types are the least employed, with a median coverage of 4.4\%.

\begin{rqsummary}{Finding \#2}
  Developers prioritize annotating function parameters and return types, indicating a focus on external API interfaces.
  The large majority of type hints rely on built-in types.
\end{rqsummary}

\subsubsection*{\textbf{\rqc}}


We detected \rqcTotalPeriods{} annotation events for all 139 repositories with at least one type hint.
Of the \rqcUniqueMembers{} type hints with a recorded annotation history, \rqcIntactN{} (\rqcIntactPct\%) remained intact after their introduction, meaning they were never changed or removed.
The remaining 84,896 type hints experienced at least one further event: \rqcModifiedN{} (\rqcModifiedPct\%) were modified at least once, and \rqcRemovedMemberN{} (\rqcRemovedMemberPct\%) were removed from the codebase at some point after their introduction.
We also observed type hints modified multiple times; when this happens, subsequent changes are performed in a 5~day interval (median).

Table~\ref{tab:rq3_member_types} presents the proportion of changes by member type.
Parameters account for the largest share (\rqcParamPct\%), followed by return types (\rqcReturnPct\%) and variables (\rqcVarPct\%).
This aligns with their annotation rates from RQ2, where parameters were the most consistently annotated---here they also lead in modification frequency.

\begin{table}[htbp]
  \centering
  \caption{Proportion of type hint changes by member type.}
  \label{tab:rq3_member_types}
  \begin{tabular}{lrr}
    \toprule
    Member Type & Count    & Proportion (\%) \\
    \midrule
    parameter   & 46{,}958 & 52.34           \\
    return      & 24{,}444 & 27.25           \\
    variable    & 18{,}317 & 20.42           \\
    \bottomrule
  \end{tabular}
\end{table}



\begin{figure}[htbp]
  \centering
  \includegraphics[width=1\linewidth]{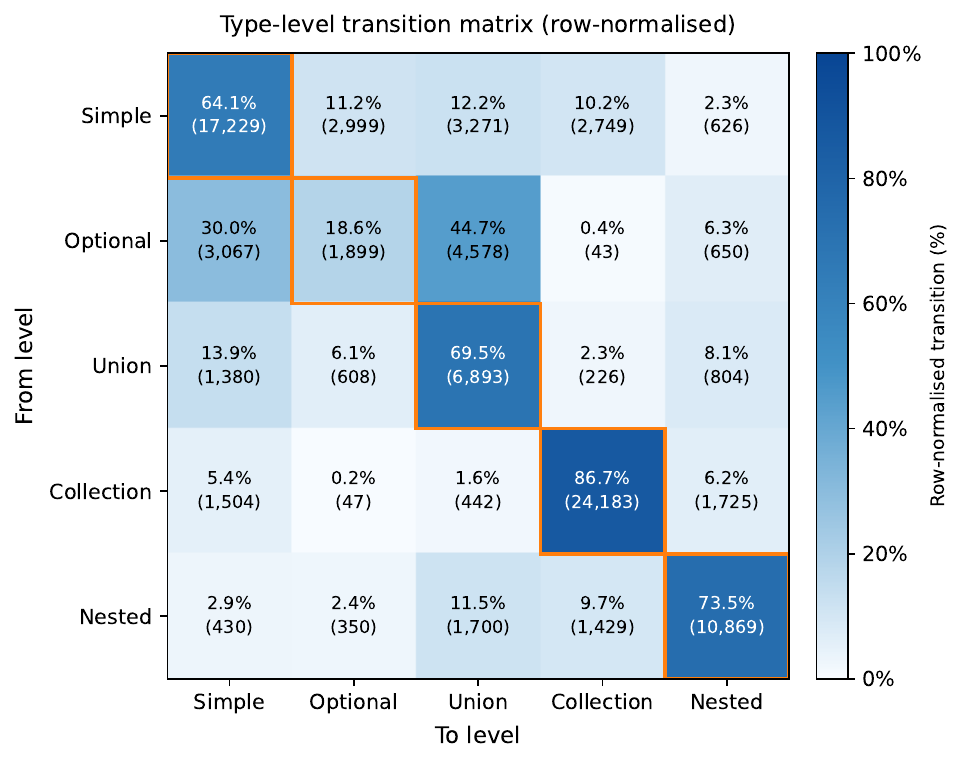}
  \caption{A heatmap showing transitions between type complexity levels. Original types are represented by rows, while target types are represented by columns.}
  \label{fig:rq3_sankey}
\end{figure}

\noindentparagraph{\textbf{Type Hint Evolution.}}
Figure~\ref{fig:rq3_sankey} illustrates how annotations migrate between type complexity levels.
\lvlUnion{} is the primary destination of type changes;
9,991 type hints migrate to \lvlUnion{} from other levels, whereas 3,018 leave it.
This makes \lvlUnion{} the complexity level with the highest incoming migrations, with \rqcUnionNetShare\% of all changes across levels.
\lvlCollection{} also receives more types than it loses, with 4,447 incoming and 3,718 outgoing transitions ($+729$).

By contrast, \lvlOptional{} is the primary source of type changes, with 4,334 type hints leaving this level for others.
Put differently, only \rqcOptionalRetention\% of \lvlOptional{} annotations remain at this level after a change.
Such relevant loss caught our attention, so we investigated some examples to understand the reasons behind this pattern.
Figure~\ref{fig:rq3_optional_union} illustrates an instance of this change.
The \texttt{date\_str} parameter in rq's \texttt{str\_to\_date} utility originally accepted optional bytes (\texttt{Optional[bytes]}), treating \texttt{None} as the absence of a date string.
When the library was extended to also accept plain string input, the annotation was updated to \texttt{Union[bytes, str]}, simultaneously broadening the input domain to include \texttt{str} and making the parameter required (removing \texttt{None}).
The commit message \textit{``Modified str\_to\_date to accept both string and bytes''} confirms this was a targeted feature extension.
\lvlSimple{} types are the second source of changes, with \rqcSimpleOutN{} (\rqcSimpleOutPct\%) leaving this level for others.
From them, \rqcSimpleToUnionN{} (\rqcSimpleToUnionPct\%) migrate to \lvlUnion, \rqcSimpleToCollectionN{} (\rqcSimpleToCollectionPct\%) to \lvlCollection, and \rqcSimpleToOptionalN{} (\rqcSimpleToOptionalPct\%) to \lvlOptional, consistent with a gradual shift toward more expressive, structured annotations over time.

\begin{figure}[htbp]
  \centering
  \begin{minted}[
        autogobble,
        linenos=false,
        fontsize=\small,
        escapeinside=||
      ]{python}
        def str_to_date(
        |\colorbox{red!12}{\makebox[0.97\linewidth][l]{\strut\textcolor{red!75!black}{-}    date\_str: Optional[bytes]}}|
        |\colorbox{green!12}{\makebox[0.97\linewidth][l]{\strut\textcolor{green!50!black}{+}    date\_str: Union[bytes, str]}}|
        ) -> Optional[datetime]: ...
    \end{minted}
  \caption{An example of \lvlOptional~$\to$~\lvlUnion{} change, extracted from rq project (commit \href{https://github.com/rq/rq/commit/2ab5ca074a594f03a24b2cd7a6752df3bf20114e\#diff-131ce6502fe39f72674a7d588e39152a3d8eea79d8ca3483b951f6957d0ce8f1}{\texttt{2ab5ca07}}).}
  \label{fig:rq3_optional_union}
\end{figure}

Finally, \lvlNested{} and \lvlUnion{} remain largely stable;
the first retains 86.7\% of its annotations after introduced, while the latter retains 73.5\%.
This suggests that once a type hint reaches a certain level of expressiveness, developers rarely simplify it back to a less expressive form.



\begin{rqsummary}{Finding \#3}
  The majority of type hints are introduced and remain unchanged (\rqcIntactPct\%).
  Once changed, type hints increasingly migrate to more expressive levels, e.g.,~\lvlOptional{} $\to$ \lvlUnion\ (\rqcOptionalToUnionPct\%).
  Parameters change most often (52.3\%), usually alongside return types and local variables.
\end{rqsummary}

\subsubsection*{\textbf{\rqd}}

\begin{figure*}[t]
  \centering
  \begin{tikzpicture}[every node/.style={align=center}]
    \tikzset{
      separator/.style={draw=black, line width=2pt},
      vbox/.style={rounded corners, dashed, gray!50},
      lbl/.style={font=\small},
      ldev/.style={blue!80!black,   font=\small\bfseries},
      lpyr/.style={orange!80!black, font=\small\bfseries},
      lnei/.style={black, font=\small},
      pct/.style={font=\normalsize\bfseries},
    }

    \begin{scope}[shift={(0,0)}]
      \node[font=\normalsize\bfseries] at (0,1.9) {Overall};
      \filldraw[fill=blue!50,   fill opacity=0.35, draw=blue!70]   (-0.45,0) circle (0.85cm);
      \filldraw[fill=orange!70, fill opacity=0.35, draw=orange!70] ( 0.45,0) circle (0.85cm);
      \node[ldev] at (-0.65, 1.1) {Dev};
      \node[lpyr] at ( 0.65, 1.1) {Pyright};
      \node[lbl] at (-0.9, 0) {\rqdDevOnly};
      \node[pct]  at (0, 0) {\rqdBothVenn};
      \node[lbl] at (0.9, 0) {\rqdPyrOnly};
      \draw[vbox] (-1.75,-1.55) rectangle (1.75,1.55);
      \node[lnei] at (0,-1.25) {\rqdNeither};
    \end{scope}

    \draw[separator] (2.1,-2.0) -- (2.1,2.0);

    \begin{scope}[shift={(4.2,0)}]
      \node[font=\normalsize\bfseries] at (0,1.9) {Parameter};
      \filldraw[fill=blue!50,   fill opacity=0.35, draw=blue!70]   (-0.45,0) circle (0.85cm);
      \filldraw[fill=orange!70, fill opacity=0.35, draw=orange!70] ( 0.45,0) circle (0.85cm);
      \node[ldev] at (-0.65, 1.1) {Dev};
      \node[lpyr] at ( 0.65, 1.1) {Pyright};
      \node[lbl] at (-0.9, 0) {\rqdParamDevOnly};
      \node[pct]  at (0, 0) {\rqdParamBoth};
      \node[lbl] at ( 0.9, 0) {\rqdParamPyrOnly};
      \draw[vbox] (-1.75,-1.55) rectangle (1.75,1.55);
      \node[lnei] at (0,-1.25) {\rqdParamNeither};
    \end{scope}

    \begin{scope}[shift={(8.4,0)}]
      \node[font=\normalsize\bfseries] at (0,1.9) {Variable};
      \filldraw[fill=blue!50,   fill opacity=0.35, draw=blue!70]   (-0.45,0) circle (0.85cm);
      \filldraw[fill=orange!70, fill opacity=0.35, draw=orange!70] ( 0.45,0) circle (0.85cm);
      \node[ldev] at (-0.65, 1.1) {Dev};
      \node[lpyr] at ( 0.65, 1.1) {Pyright};
      \node[lbl] at (-0.9, 0) {\rqdVarDevOnly};
      \node[pct]  at (0, 0) {\rqdVarBoth};
      \node[lbl] at ( 0.9, 0) {\rqdVarPyrOnly};
      \draw[vbox] (-1.75,-1.55) rectangle (1.75,1.55);
      \node[lnei] at (0,-1.25) {\rqdVarNeither};
    \end{scope}

    \begin{scope}[shift={(12.6,0)}]
      \node[font=\normalsize\bfseries] at (0,1.9) {Return};
      \filldraw[fill=blue!50,   fill opacity=0.35, draw=blue!70]   (-0.45,0) circle (0.85cm);
      \filldraw[fill=orange!70, fill opacity=0.35, draw=orange!70] ( 0.45,0) circle (0.85cm);
      \node[ldev] at (-0.65, 1.1) {Dev};
      \node[lpyr] at ( 0.65, 1.1) {Pyright};
      \node[lbl] at (-0.9, 0) {\rqdReturnDevOnly};
      \node[pct]  at (0, 0) {\rqdReturnBoth};
      \node[lbl] at ( 0.9, 0) {\rqdReturnPyrOnly};
      \draw[vbox] (-1.75,-1.55) rectangle (1.75,1.55);
      \node[lnei] at (0,-1.25) {\rqdReturnNeither};
    \end{scope}

  \end{tikzpicture}
  \caption{Proportion of members covered by developer annotations only, Pyright inference only, both (bold), or neither---overall and by member type.}
  \label{fig:rq4_venn}
\end{figure*}
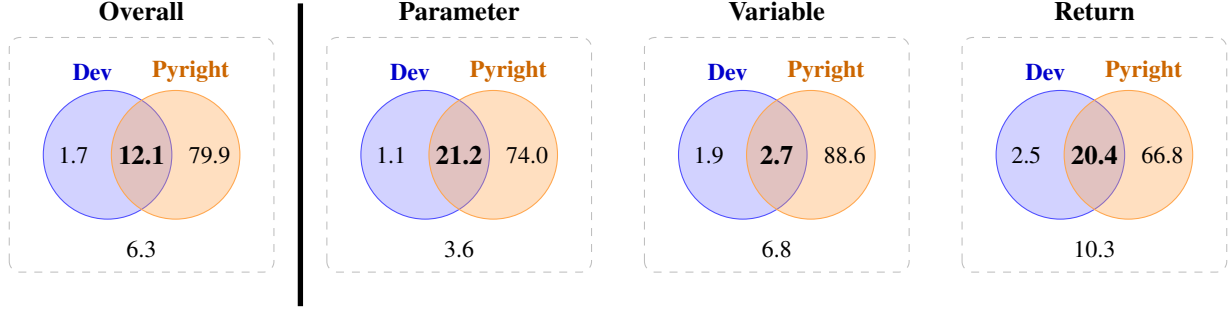

Figure \ref{fig:rq4_venn} depicts four Venn diagrams. Overall and one for each member type showing the proportion of members that are (a) annotated with type hints (in blue), (b) inferred by Pyright (in orange), (c) covered by both (intersection), and (d) ignored by both (outside both sets).
In general, \rqdNeither\% of eligible members are neither annotated nor inferred.
Pyright is able to infer types for \rqdUnannotatedInferable\% of unannotated members, while type hints account for only \rqdDevOnly\% of members that Pyright cannot infer.

When we break down the coverage by member type, we observe that \textit{parameters} and \textit{return types} have a meaningful overlap (\rqdParamBoth\% and \rqdReturnBoth\%, respectively), suggesting that developers annotate even where Pyright can infer the type.
For \textit{return types}, however, the dominant zone is \textit{neither} (\rqdReturnNeither\%): Pyright cannot infer most return types from the function body alone, so developer annotations are essential here.
By contrast, \textit{variables} are dominated by Pyright-only coverage (\rqdVarPyrOnly\%), indicating that developers rarely annotate local variables.
We believe two reasons explain this particularity:
(a) Pyright can infer local variable types directly from their assignment expressions; and
(b) developers can infer the eligible types more easily from the surrounding code.

We also investigated divergences between type hint annotations and types inferred by Pyright.
From the \rqdBothCount{} members where both sources provide a type, Table~\ref{tab:rq4_funnel} breaks down agreement by complexity level; each row shows the distribution of Pyright-inferred levels for a given developer type hint.
\lvlSimple, \lvlUnion, and \lvlCollection{} annotations show strong agreement with Pyright (\rqdSimpleAgree\%, \rqdUnionAgree\%, and \rqdCollectionAgree\%, respectively).
By contrast, \lvlOptional{} annotations almost never match Pyright (0.0\%): Pyright infers the semantically equivalent \lvlUnion{} type (e.g.,~\texttt{int\,|\,None}) in \rqdOptionalToUnion\% of cases, revealing a systematic syntactic preference where developers favor the more concise \texttt{Optional[T]} form.
\lvlNested{} annotations also show lower agreement (\rqdNestedAgree\%), with Pyright preferring \lvlUnion{} in \rqdNestedToUnion\% of cases, suggesting that developers sometimes use more structured types than Pyright would infer.


\begin{table}[ht]
  \centering
  \caption{Distribution of Pyright-inferred complexity levels per developer type hint level (row \%). Each row sums to 100\%; bold values are exact level agreements.}
  \label{tab:rq4_funnel}
  \begin{tabular}{lrrrr}
    \toprule
    \multirow{2}{*}{\textbf{Type Hint}} & \multicolumn{4}{c}{\textbf{Pyright}}                                                                                          \\
    \cmidrule(l){2-5}
                                        & \textbf{Simple}                      & \textbf{Union}               & \textbf{Collection}          & \textbf{Nested}          \\
    \midrule
    \textbf{Simple}                     & \textbf{\rqdSimpleAgree}             & \rqdSimpleToUnion            & \rqdSimpleToCollection       & \rqdSimpleToNested       \\
    \textbf{Optional}                   & \rqdOptionalToSimple                 & \textbf{\rqdOptionalToUnion} & \rqdOptionalToCollection     & 0.0                      \\
    \textbf{Union}                      & \rqdUnionToSimple                    & \textbf{\rqdUnionAgree}      & \rqdUnionToCollection        & \rqdUnionToNested        \\
    \textbf{Collection}                 & \rqdCollectionToSimple               & \rqdCollectionToUnion        & \textbf{\rqdCollectionAgree} & \rqdCollectionToNested   \\
    \textbf{Nested}                     & \rqdNestedToSimple                   & \rqdNestedToUnion            & \rqdNestedToCollection       & \textbf{\rqdNestedAgree} \\
    \bottomrule
  \end{tabular}
\end{table}

\begin{rqsummary}{Finding \#4}
  Only \rqdNeither\% remain untyped even after combining developer annotations and Pyright inference, with return types being the most affected.
  Developers tend to annotate parameters and return types, regardless of Pyright's ability to infer them.
  When such override happens, developers often simplify inferred types.
\end{rqsummary}

\section{Discussion}
\label{sec:discussion}

This study provides an empirical view of how type hints are adopted, used, maintained, and complemented by static inference in Python libraries and frameworks. The results show that type hints have become a common feature in popular Python components, but their use remains selective rather than systematic. Most libraries and frameworks contain at least one type hint, yet the median coverage remains low. Type hints are present in most projects, but they are not yet consistently maintained as an engineering layer across the whole codebase.

\noindentparagraph{Typehints are API-Oriented.}
The most important insight is that type hints in libraries and frameworks appear to be primarily API-oriented. Developers annotate parameters and return types more frequently than variables, and these two categories evolve together. This pattern indicates that maintainers tend to use type hints where they have the highest communicative and contractual value: at the boundary between the library and its users. For libraries and frameworks, this is a significant finding because the public API is not only an internal design artifact; it is the main point of interaction with downstream developers, IDEs, static analyzers, documentation generators, and automated tools. Type hints, therefore, operate as an ecosystem-facing artifact rather than merely local code metadata.

This API-oriented use of type hints has direct implications for practitioners. Library maintainers do not necessarily need to pursue full type coverage as an immediate goal. Instead, our results suggest that a pragmatic typing strategy may focus first on public-facing members: exported functions, class constructors, method parameters, return values, and extension points used by clients. This strategy aligns with the observed behaviour of real-world projects and may offer a better cost-benefit trade-off than attempting to annotate every eligible member. For maintainers working with large legacy codebases, incremental typing policies could therefore start at the API boundary and gradually move inward as the code evolves.

For researchers, this finding suggests that future studies should move beyond global annotation coverage. Counting all eligible members equally provides a useful first approximation, but it may obscure the real engineering value of type hints in libraries and frameworks. A project with low global coverage may still provide strong typing support for its public API. In contrast, another project with higher global coverage may annotate mostly internal or low-impact members. Future empirical work should therefore distinguish public API coverage, internal implementation coverage, test coverage, and generated-code coverage. Such distinctions would allow researchers to evaluate not only how much typing there is, but also where typing matters most.

For practitioners, the findings support an incremental and pragmatic typing strategy. The evidence does not suggest that real-world library maintainers necessarily pursue full type coverage. Instead, they concentrate annotations around function signatures. This evidence provides a practical path for maintainers of large Python codebases: prioritize exported functions, public classes, constructors, callback interfaces, extension points, and return values before annotating every local variable. Such a strategy aligns with observed practice and may provide the best return on effort by improving API comprehension, client-side static analysis, IDE support, and documentation quality.

\noindentparagraph{Maintenance Practices on Type Hints.}
Our study shows that annotations are introduced, changed, and removed over time, with a substantial number of changes affecting parameters and return types. It means that type hints evolve as the software's design evolves.
These findings have direct implications for library maintenance.
Engineers should treat type hints as part of a library's public contract.
Therefore, when a parameter annotation or return annotation changes, the change may affect downstream users in ways similar to a behavioural or signature-level API change.
Maintainers should review type changes carefully, document them in release notes when they affect public APIs, and include type-checking in continuous integration.
In mature projects, annotation changes should be considered in compatibility policies, especially when they narrow accepted inputs or change the apparent return contract of a function. The maintenance findings reinforce this point. Type hints are not static decorations added once and forgotten.

The observed movement toward \texttt{Union} types is also noteworthy.
On one hand, it may indicate that developers progressively refine annotations as they discover more precise behavioural contracts.
On the other hand, some transitions may reflect syntactic modernization in the Python type system rather than a true semantic change.
For example, moving from \texttt{Optional[T]} to \texttt{T | None} is different from broadening a parameter from \texttt{Optional[int]} to \texttt{Union[int, float, None]}.
Future work should separate syntactic migration from semantic evolution.

\noindentparagraph{Opportunities for tool builders.}
For tool engineers, the results expose a clear opportunity: current tooling should support selective, interface-first, and evolution-aware typing. Since developers already prioritize parameters and return types, annotation recommendation tools should focus first on public signatures rather than indiscriminately suggesting annotations for all members. Tools that recommend high-impact annotations could rank candidates based on API visibility, call frequency, documentation presence, downstream usage, or the likelihood of type-checking benefits.

The disagreement between Pyright inference and developer annotations is also instructive. Some disagreements may be erroneous, but others may reflect intentional abstraction. A static analyzer may infer a concrete implementation type, while the developer intentionally writes a broader type to preserve substitutability, hide implementation details, or document the intended interface. Tool engineers should therefore avoid treating all mismatches as defects. More useful tools would classify mismatches such as semantic incompatibility, equivalent syntax, intentional generalization, possible over-specific annotation, and possible under-specific annotation.

\noindentparagraph{Gradual Typing involves tradeoffs.}
Type hints should not be presented only as optional syntax or as a way to make Python resemble statically typed languages.
Their practical role is broader: they document expectations, support tool-based reasoning, improve IDE feedback, help detect inconsistencies, and communicate API contracts to users.
Therefore, developers must learn to balance the benefits of type hints against their maintenance overhead.

Such mindset should be reflected in teaching.
Students do not need to learn that "everything must be typed" as a universal rule.
A more realistic lesson is that typing decisions involve trade-offs.
Instructors can ask students to prioritize annotations for public interfaces, justify why certain internal variables remain unannotated, and evaluate how annotations affect readability, maintainability, and tool feedback. In testing assignments, students can combine type checking with unit tests and documentation checks. In software architecture or framework-development courses, type hints can be discussed as lightweight interface contracts that support evolution and reuse.

\section{Threats to Validity}
\label{sec:threats}

We discuss the main threats to validity of this study following the classification by Wohlin et al.~\cite{Wohlin2012}.

\subsection*{Construct Validity}

\noindentparagraph{Type hint coverage as a proxy.}
The \textit{type hint coverage} metric counts annotated members as a proportion of eligible members.
It treats all members equally, regardless of their public visibility, centrality to the API, or semantic importance.
A repository may score low coverage while having all its public API surface fully annotated, or score high by annotating only trivial internal variables.

\noindentparagraph{Pyright as a proxy for inference capability.}
For RQ4, we use Pyright as the static type checker for performing inference.
We chose Pyright because it offers the highest known adherence to the official Python typing specification among mature tools ($\approx$98\%, versus $\approx$53\% for mypy~\cite{PyrightConformance2024}), and because it underpins Pylance, the default language server shipped with VS Code’s Python extension, making it the most widely used inference engine in real-world practice~\cite{Pylance2020}.

\noindentparagraph{Arbitrary threshold.}
The 10\% overall coverage cutoff used to select \libs\ with ``systematic'' type hint usage in RQ2 is a pragmatic choice rather than a theoretically grounded one.
Repositories just below this threshold are excluded from the analysis, which may slightly underestimate adoption.

\subsection*{Internal Validity}

\noindentparagraph{Git history assumptions.}
The annotation history analysis (RQ3) replays each file's commit log sequentially.
Non-linear histories introduced by rebases, squash merges, or force-pushes may cause the tool to miss intermediate annotation states or to attribute changes to the wrong commits.
Shallow clones would also break this procedure entirely; we mitigated this by cloning the full history of each repository.

\subsection*{External Validity}

\noindentparagraph{Repository selection.}
Our dataset consists of the top-1,000 Python repositories on GitHub ranked by star count.
Popular, starred projects tend to be better maintained and more likely to adopt modern language features such as type hints.
Results may not generalize to less popular, internal, or proprietary Python libraries.

\noindentparagraph{Platform and language restriction.}
We studied only GitHub-hosted, English-language, open-source repositories.
Repositories hosted on other platforms (GitLab, Bitbucket) or primarily documented in other languages were excluded.
This limits generalizability to the broader Python ecosystem.

\subsection*{Conclusion Validity}

\noindentparagraph{Statistical choices.}
We use Spearman's rank correlation to assess statistical significance.
Different choices of non-parametric tests could yield different results.
We mitigated this by reporting effect sizes and $p$-values alongside all correlation results, and by selecting thresholds consistent with prior empirical software engineering work.

\noindentparagraph{Snapshot in time.}
The dataset reflects the state of each repository at a specific point in time.
The Python type hint ecosystem is actively evolving---new PEPs, type checker improvements, and community tooling may shift adoption patterns after our collection date.

\section{Related Work}
\label{sec:related-work}


\subsection{Type Systems and Software Quality}

Empirical research shows the advantage of using type systems to detect errors before runtime, to reduce debugging costs, and to improve code maintainability~\cite{Gao2017,Hanenberg2014,Bogner2022}.
\citet{Hanenberg2014} have shown through a controlled experiment with 33 subjects that static typing assists developers in better navigating through the code base, reducing the effort to maintain software projects.
\citet{Gao2017} found that static type checkers implemented in Flow and TypeScript the supersets of the JavaScript language, could detect approximately 15\% of real-world bugs in JavaScript projects.
In a similar study, \citet{Bogner2022} has shown that TypeScript applications present better code quality and legibility when compared to Java\-Script ones.
\citet{Mezzetti2018} showed that many breaking changes detected in JavaScript projects are due to type-related issues.

\subsection{Gradual Typing and Adoption}

The theoretical foundation for gradually typed languages was established by \citet{Siek2007}, who introduced the concept of \textit{gradual type system} that seamlessly combines statically and dynamically typed code through a consistency relation between types. Python's type hint mechanism, introduced via PEP~484, draws on this framework by making annotations optional and delegating enforcement to external static analysis tools such as mypy and pyright, rather than the interpreter itself.

Prior work has also investigated the adoption of optional and gradual type systems~\cite{DiGrazia2022,Scarsbrook2023}.
\citet{DiGrazia2022} conducted a large-scale empirical study on type annotation evolution in Python, where they analyzed 1.4 million annotation changes extracted from 9,655 popular GitHub projects.
The authors found that type hints are increasingly adopted and, once introduced, can help developers detect additional type errors.
\citet{Scarsbrook2023} investigated TypeScript adoption among 454 repositories, finding that while the TypeScript compiler is rapidly being adopted, the adoption of language-specific features varies significantly between the analyzed projects.

\subsection{Python Type Annotation Practices}

\citet{Mir2021} presents \textit{ManyTypes4Py}, a dataset with approximately 870K type annotations from Python projects, instrumented for training machine learning models specialised in performing type inference.
Several studies have focused specifically on Python type annotation practices.
\citet{Jin2021} conducted a large-scale study of 2,862 Python projects to examine \textit{where} developers start annotating and what factors influence adoption, finding that function parameters are consistently annotated before return types and that project size and contributor count are associated with higher annotation rates.
\citet{Lin2023} performed a broad empirical analysis of Python static type annotations across thousands of repositories, characterising annotation density, coverage growth trends, and the prevalence of different annotation constructs over time.

\textit{Unlike prior large-scale studies on Python type annotations, our work focuses on libraries and frameworks, i.e.,~the building blocks of modern software development.}

\subsection{Automated Type Inference}

A complementary line of work investigates automated approaches to infer or recommend type annotations, reducing the manual effort required from developers.
\citet{Mir2021} released the ManyTypes4Py dataset to support training of ML-based type inference models; subsequent work has produced models such as Type4Py~\cite{Mir2022} and TypeWriter~\cite{Pradel2020}, which use deep learning to predict likely type annotations from source code and usage patterns.
These tools often rely on well-annotated libraries as a source of type information, which underscores the importance of understanding annotation practices in library code---the primary focus of our study.
Improving annotation coverage and consistency in libraries would directly benefit the accuracy and applicability of such automated tools.

\section{Conclusion} 
\label{sec:conclusion}

In this work, we present an empirical study on the adoption, usage, and maintenance of type hints in popular Python libraries and frameworks.
We extracted and analyzed 649,099 type annotations declared in 152 Python \libs\ hosted on GitHub, covering 5,030,649 source code members.
While 9 out of 10 libraries use type hints at least once, half of them annotate at most 13.6\% of their members.
Maintainers focus on annotating function parameters and return types, mainly using built-in types, indicating a focus on API contracts over internal implementation details.
After tracking the timeline of \rqcTotalPeriods{} annotation modifications, we found that type hints are predominantly introduced rather than changed or removed, and that \lvlUnion{} types are the dominant destination of annotation changes, receiving 3.3$\times$ more inflows than outflows.
Finally, comparing developer annotations against Pyright's inference reveals that developers annotate precisely where automated inference is insufficient: only \rqdDevOnly\% of members are annotated by developers alone, while \rqdPyrOnly\% are covered only by Pyright. When developers do provide annotations, they systematically prefer \texttt{Optional[T]} where Pyright would produce the equivalent union syntax.

\noindentparagraph{Future Work.}
This study opens several directions for further investigation.
A qualitative analysis---e.g.,~through commit message mining or developer interviews---is needed to understand \textit{why} maintainers choose to annotate specific members and which types to use.
Regarding maintenance, future work could study how type hint changes relate to API breaking changes, deprecation cycles, or semantic versioning decisions.
A particularly promising direction is to compare the behaviour of multiple static inference tools, such as mypy, pytype, and the emerging Pyrefly~\cite{MetaTypingSurvey2024}, to assess whether the gap between developer annotations and automated inference is tool-specific or a fundamental limitation of current static analysis.
Finally, as annotation coverage grows, new automated tools could leverage our findings to recommend which members to prioritize for annotation and which type constructs are most appropriate for a given API surface.

\noindentparagraph{Replication Package.}
The dataset and scripts used in this study are publicly available at: \url{https://zenodo.org/records/21563590}.




\bibliographystyle{elsarticle-num-names}
\bibliography{main}

\end{document}